\documentclass[twocolumn]{article}
\usepackage{pdfsync}
\usepackage{hyperref}
\usepackage{verbatim}
\usepackage{graphicx}
\usepackage[english]{babel}
\usepackage[a4paper]{geometry}
\usepackage{stmaryrd}
\usepackage{named,bm,amsmath}
\parskip=\smallskipamount
\advance\textwidth by 3cm
\advance\oddsidemargin by -1.5cm
\advance\evensidemargin by -1.5cm
\advance\textheight by 3cm
\advance\topmargin by -1.5cm

\def\..{\,\mathpunct{\ldotp\ldotp}} 

\renewcommand{\epsilon}{\varepsilon}

\title{Fibonacci Binning}
\author{Sebastiano Vigna\footnote{The author has been supported by the EU-FET grant
NADINE (GA 288956).}\\\texttt{vigna@acm.org}}
\begin{document}
\bibliographystyle{named}

\maketitle

\begin{abstract}
This note argues that when dot-plotting 
distributions typically found in papers about web and social networks
(degree distributions, component-size distributions, etc.), and more generally
distributions that have high variability in their tail, an exponentially binned version 
should always be plotted, too, and suggests \emph{Fibonacci binning} 
as a visually appealing, easy-to-use and practical choice.
\end{abstract}

\section{Introduction}

\begin{figure}
\centering
\includegraphics[scale=.6]{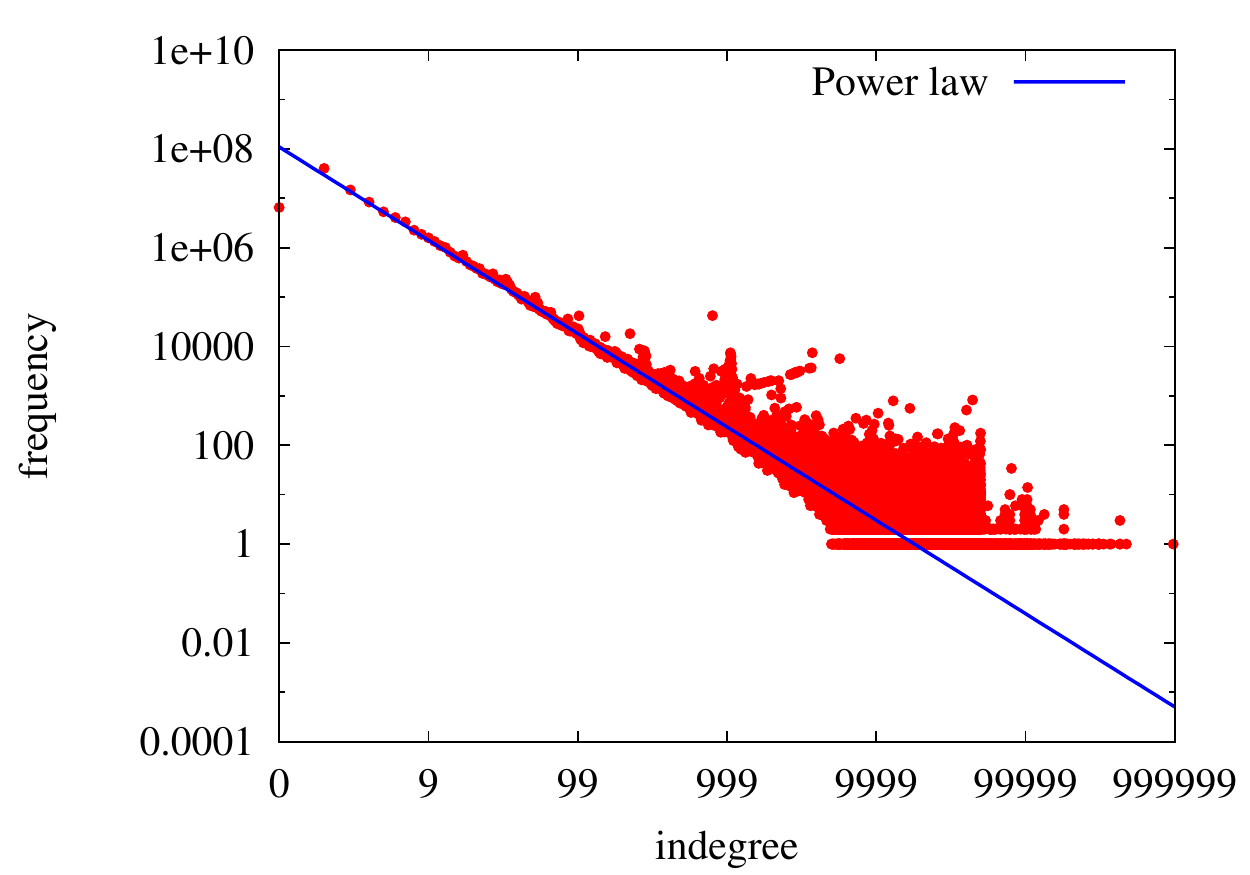}
\caption{\label{fig:in}The frequency dot plot of the indegree distribution  of
a 106 million pages snapshot of the \texttt{.uk} web crawled in May 2007
(available at \texttt{http://law.di.unimi.it/}). The line shows a power law with exponent $1.89$.}
\end{figure}

The literature about web and social networks has been in the last decade
literally inundated by dot plots like Figure~\ref{fig:in}: for each abscissa $x$
(usually, a degree or a size), a dot is plotted at coordinates $\langle
x,y\rangle$, where $y$ the frequency of the element (nodes, components) with
feature $x$.

Misuses of such graphs have been abundantly described
elsewhere~\cite{WADMI,LADTTSFG}---in particular, their role in convincing people
easily that some \emph{power law} would fit the empirical data distribution just
by plotting lines through the ``cloud of points'' instead of using some statistically
sound test.\footnote{Interestingly, the
same considerations appear to have been common knowledge at least a decade ago
in other areas~\cite{HerSOCES}.}

The main problem of such plots is that the tail is, actually, unfathomable: due
to the sparsity and high variability of the points in the right part of the
graph, it is impossible to infer visually anything about the behavior of the
tail of the distribution.

A sound solution is using a standard statistical methodology as discussed in detail, for instance,
in~\cite{CSNPLDED}: first finding the starting point by max-likelihood estimation, then
computing a $p$-value, and finally comparing with other models. Nonetheless, visual inspection of
plots remains useful to get a ``gut feeling'' of the behavior of the distribution.

One alternative suggested in~\cite{LADTTSFG} is using \emph{size-rank plots}---the 
numerosity-based discrete analog of the complementary
cumulative distribution function in probability.\footnote{Limitations of size-rank
plots are discussed in~\cite{HerSOCES}.} To each abscissa $x$ we
associate the sum of the frequencies of all data points with abscissa greater
than or equal to $x$. The plot we now obtain is monotonically decreasing, there
is no cloud of points, and the shape of the tail will be a straight line if and
only if the distribution is a power law. 

The main problem is that people \emph{love} frequency dot plots, and it should be relatively easier
to convince them to apply a binning (which, among other things, looks nice) than
change the type of diagram altogether.

\section{Fibonacci binning}

Fibonacci binning is a simple exponential (or logarithmic, depending on the viewpoint) 
discrete binning technique: bins are sized like the Fibonacci numbers.
It displays nicely on a log-log scale because Fibonacci numbers are multiplicatively spaced  
approximately like the golden ratio,
and it has the useful feature that the first two bins are actually data points. This feature comes very handy as
most empirical distributions found in web and social networks have slightly different behavior on the first one or 
two data points.\footnote{This issue is actually solved in most papers by \emph{not} plotting the value for abscissa zero, which
happens automatically if you choose to plot in log-log scale in any plotting package known to the author.}
Moreover, Fibonacci binning is less coarse than the common power-of-$b$ binnings
(e.g., $b=2,10$), which should make the visual representation more accurate~\cite{ViCPLDBED}. 

\begin{figure}
\centering
\includegraphics[scale=.6]{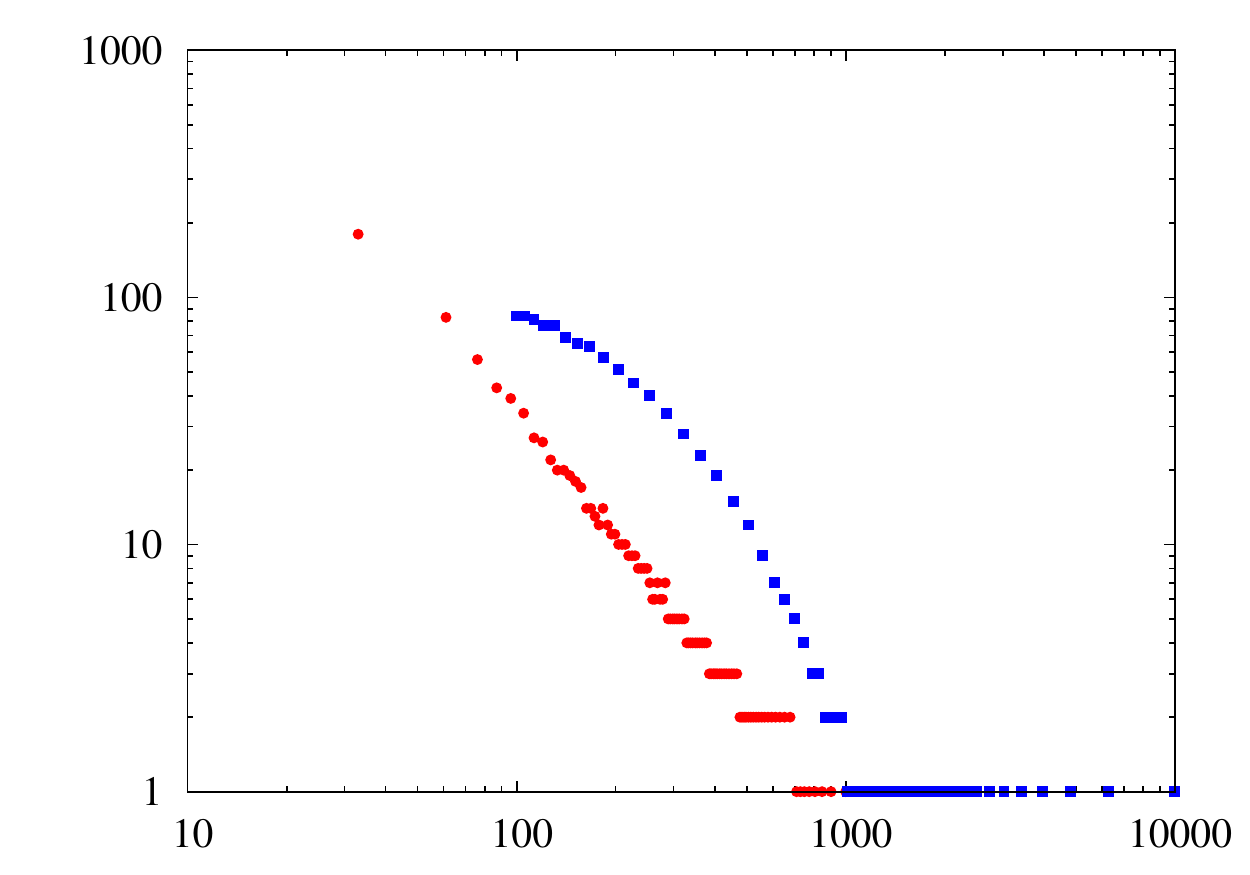}
\caption{\label{fig:fake}A dot plot of two distribution described in~~\protect\cite{LADTTSFG}. 
Can you guess which one comes from a power law?}
\end{figure}

Binning is essential for getting a graphical understanding of the tail of dot
plots.\footnote{Exponential binning is discussed in detail in~\cite{MilPLDIS}, where the authors suggest it
as a better way to fit power laws, even with respect to size-rank plots.} 
Consider, for instance, the famous pathological example shown in
Figure~\ref{fig:fake}, which is discussed in~\cite{LADTTSFG}.\footnote{The code
to generate the pathological example can be found at
\url{http://hot.caltech.edu/topology/RankVsFreq.m}.} The figure shows two
typical frequency dot plots. The (obvious) reason the example is pathological is that the
plot looking like a straight line is a sample from an exponential distribution, whereas
the curved plot is a sample from a power-law distribution.
Of course, you are supposed to think the exact contrary, and if you've seen many
dot plots like Figure~\ref{fig:in} some reasonable doubts about ``visual
distribution fitting'' using frequency plots should surface to your mind.

Binning, however, comes to help.
By averaging the values across a contiguous segment of abscissas, we obtain a more regular set of points
(essentially, the midpoints of the histograms on the same intervals) that we can connect 
to get more insight on the actual
shape of the curve. Note that the lines connecting the point are absolutely imaginary; they're just a visual clue---they are
not part of the data.

\emph{Kernel density estimation} is another technique widely used for this purpose, but it
does not really work well with discrete distributions and in particular with distributions with a ``starting point''.

More in detail, let $F_0=1$, $F_1=1$, $F_n=F_{n-1}+F_{n-2}$,\footnote{It is also customary to use $F_0=0$, $F_1=1$ as initial condition
for the Fibonacci numbers, but
our choice makes the following notation slightly easier to read.} assume that we have a 
\emph{starting offset} $s$ (usually $0$ or $1$) and data $\langle x_i,y_i\rangle$ with $x_i$ distinct integers satisfying $x_i\geq s$.
The binning intervals $[\ell_j\..r_j)$, $j\geq 0$, are then built starting at $s$ using lengths 
$F_0$,~$F_1$,~$F_2$, $\dots\,$:
\begin{align*}
[\ell_0\..r_0) &= [s + F_1 - 1\..s + F_2 - 1) \\
[\ell_1\..r_1) &= [s + F_2 - 1\..s + F_3 - 1) \\
               & \dots\\
[\ell_j\..r_j) &= [s + F_{j+1} - 1\..s + F_{j+2} - 1) \\
               & \dots\\
\end{align*}
Note that $r_k-\ell_k=F_k$, and that if $s=1$ the extremes of the intervals are exactly consecutive Fibonacci numbers.
The resulting binned sequence $\langle p_k,m_k\rangle$, $k\geq 0$ is
\[
\langle p_k,m_k\rangle = \Biggl\langle \ell_k + \frac{F_k-1}2, \frac1{F_k}\sum_{x_i\in[\ell_k\..r_k)}y_i \Biggr\rangle. 
\]

\begin{figure}
\centering
\includegraphics[scale=.6]{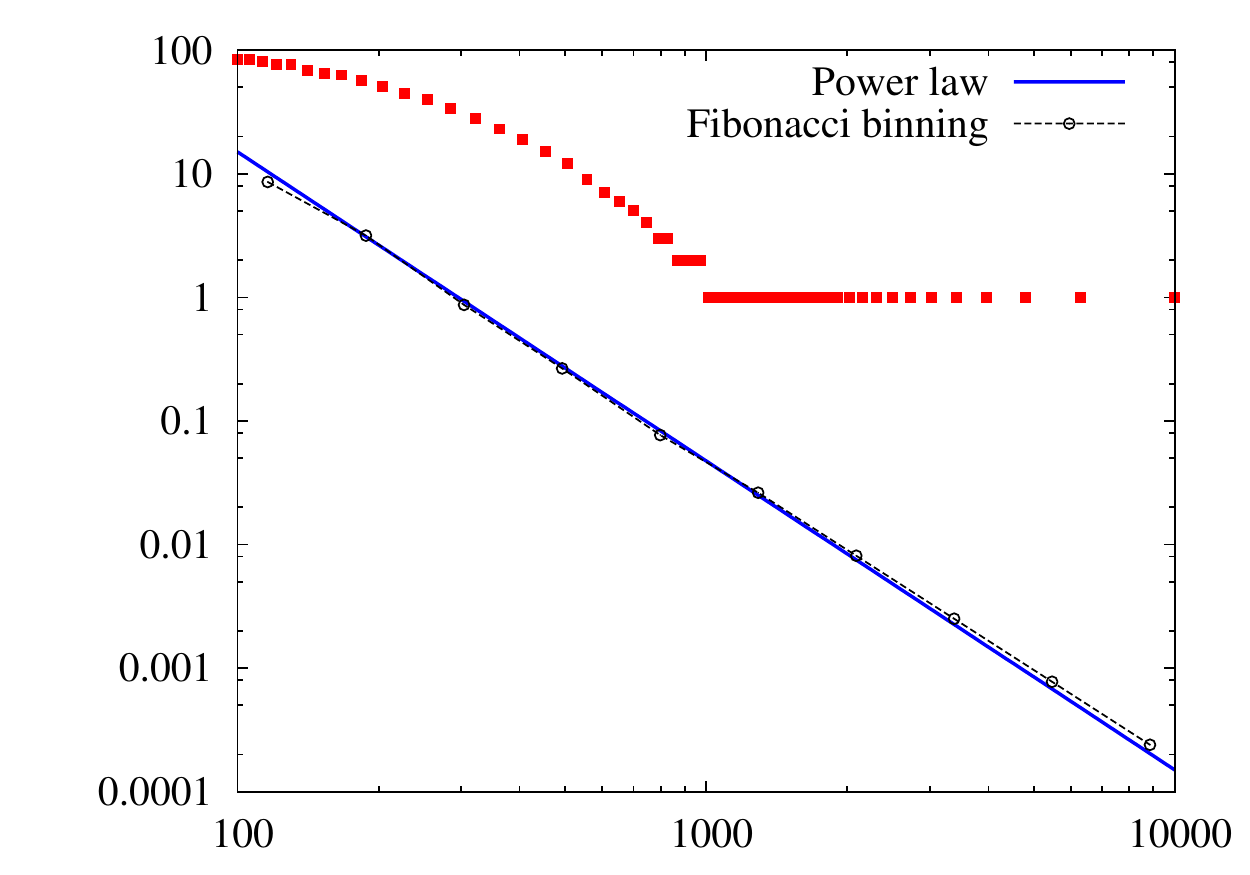}
\caption{\label{fig:pl}A dot plot of the pathological sample from a power-law distribution from~\protect\cite{LADTTSFG}, its Fibonacci binning and
the original power-law distribution used to generate the sample.}
\end{figure}

\begin{figure}
\centering
\includegraphics[scale=.6]{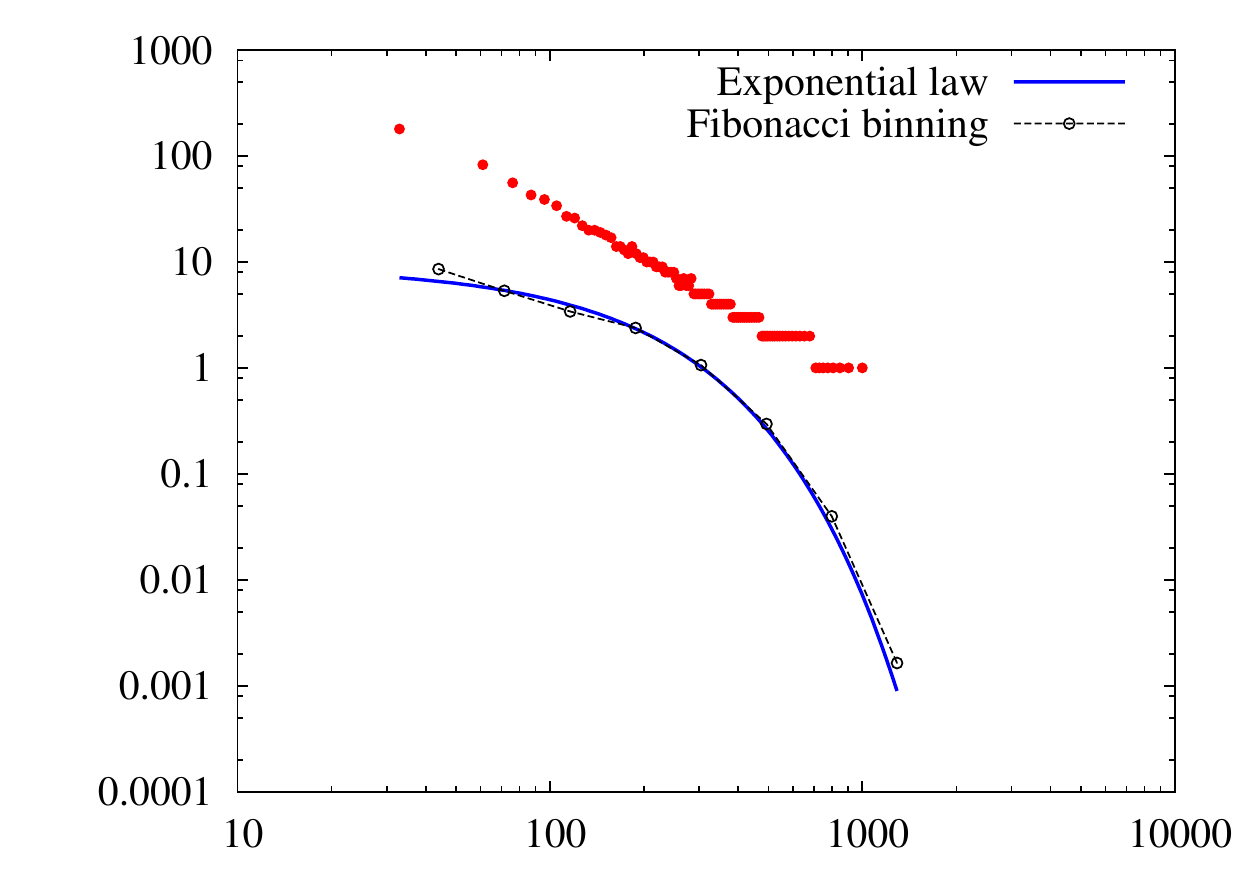}
\caption{\label{fig:exp}A dot plot of the pathological sample from an exponential distribution from~\protect\cite{LADTTSFG}, its Fibonacci binning and
the original exponential distribution used to generate the sample.}
\end{figure}

Figures~\ref{fig:pl} and~\ref{fig:exp} show the result of Fibonacci binning on
the pathological curves: the truth is easily revealed, and we obtain a very close fit with the 
distribution used to generate the sample. 

We remark that, in fact,
the pathological power-law curve is not so pathological:
\texttt{plfit}\footnote{\url{https://github.com/ntamas/plfit}} provides a best
max-likelihood fitting starting at 100 with $\alpha=2.59$ and a $p$-value
$0.154\pm0.01$, thus essentially recovering the original distribution, which has exponent $2.5$.

\begin{figure}
\centering
\includegraphics[scale=.6]{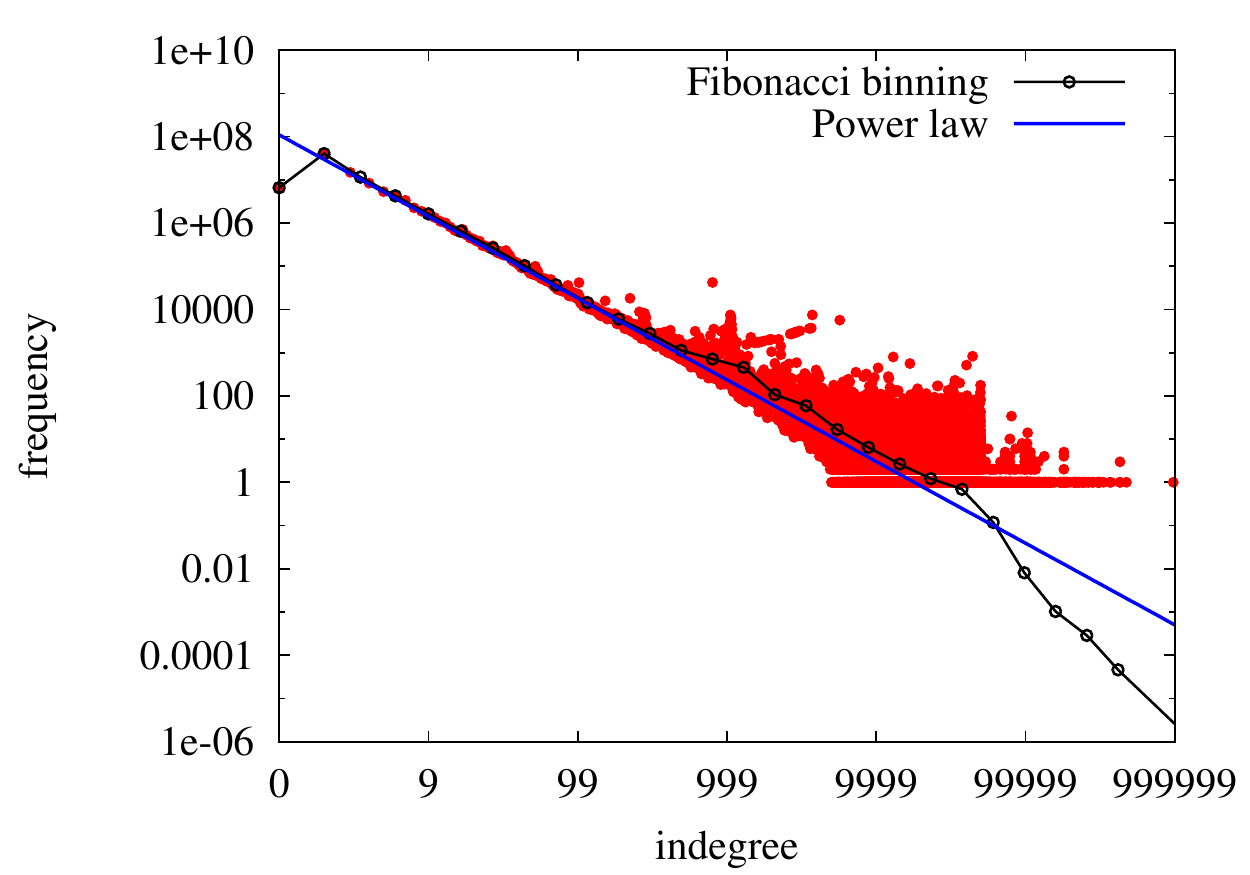}
\caption{\label{fig:in2}The plot of Figure~\ref{fig:in} with an overlapped Fibonacci binning, displaying previously undetectable concavity.}
\end{figure}

\begin{figure}
\centering
\includegraphics[scale=.6]{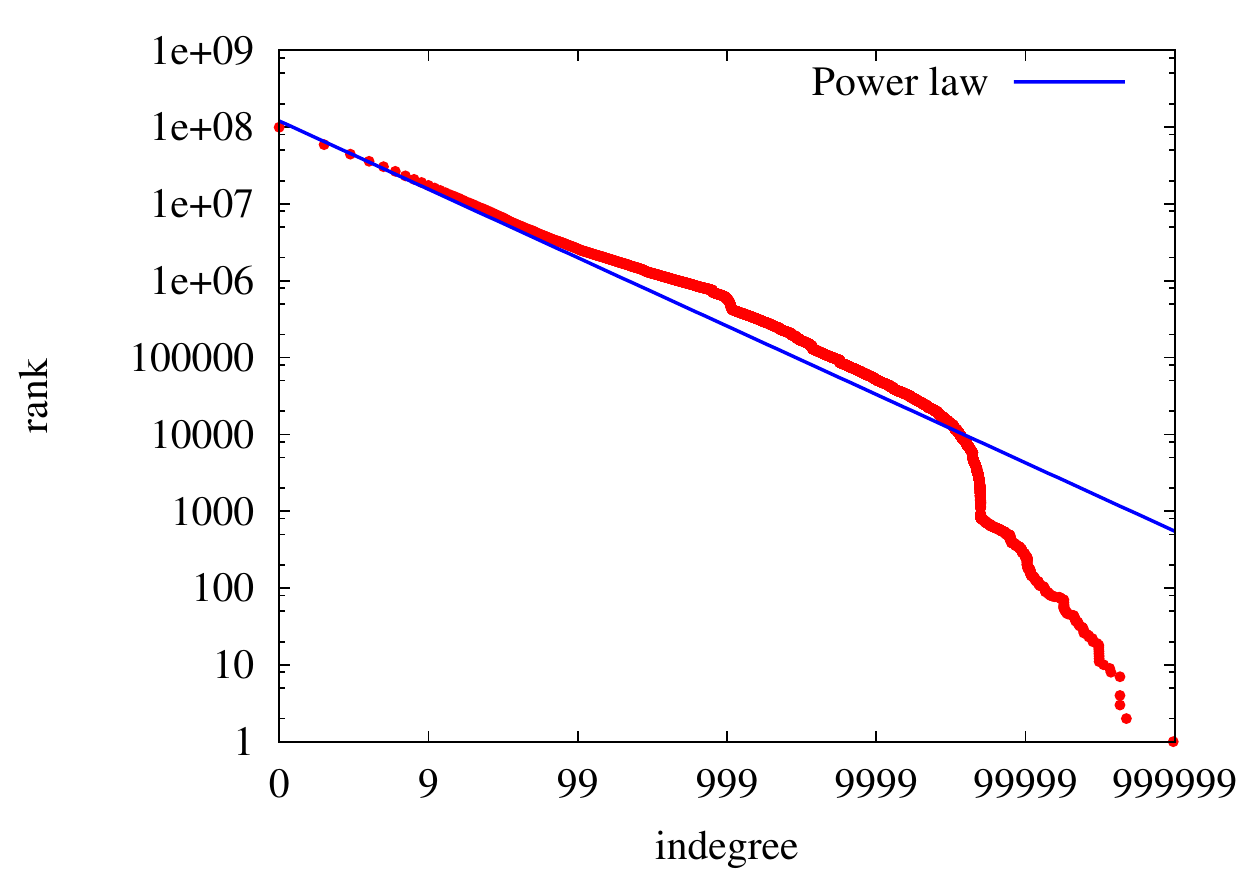}
\caption{\label{fig:ir}The size-rank plot of the data displayed in Figure~\ref{fig:in}, showing a clear concavity.}
\end{figure}

Getting back to our motivating example (Figure~\ref{fig:in}), Figure~\ref{fig:in2} shows the same data with an overlapped
Fibonacci binning, and Figure~\ref{fig:ir} shows the associated size-rank plot. The apparent fitting
of the power law is now clearly revealed as an artifact of the frequency plot, and the change
of slope actually makes unlikely the existence of a fat tail. Not surprisingly, trying to
fit a power law with \texttt{plfit} gives a $p$-value of $0\pm0.01$.

\section{Conclusions}

We hope to have convinced the reader of the advantages of Fibonacci binning. While cumulative
plots remain a somewhat more reliable and principled visual tool, and proper statistical testing
is irreplaceable, frequency plots are here to stay
and Fibonacci binning can help to make some sense out of them.

A Ruby script that computes the Fibonacci binning of a list of values is
available from the author.\footnote{\url{http://vigna.di.unimi.it/fbin.rb}}
The site of the Laboratory for Web Algorithmics\footnote{\url{http://law.di.unimi.it/}} provides examples of
frequency plot with Fibonacci binning and size-rank plots for dozens of 
networks, ranking from Wikipedia to web snapshots; it is a good place to have a taste of the
visual results.

\bibliography{biblio}

\end{document}